\documentclass[pdflatex,sn-basic]{sn-jnl}%
\usepackage{graphicx}%
\usepackage{hhline,multirow}
\usepackage{makecell}
\usepackage{amsmath,amssymb,amsfonts}%
\usepackage{booktabs}%
\usepackage{algorithm}%
\usepackage{algorithmicx}%
\usepackage{listings}%
\usepackage{tabularx}

\begin{document}

\title[Title]{Transformers-based architectures for stroke segmentation: A review}

\author[1]{\fnm{Yalda} \sur{Zafari-Ghadim}}
\author[2]{\fnm{Essam A. } \sur{Rashed}}
\author*[1]{\fnm{Mohamed} \sur{Mabrok}}\email{m.a.mabrok@gmail.com}

\affil[1]{\orgdiv{Department of Mathematics and Statistics}, \orgname{Qatar University}, \orgaddress{ \city{Doha}, \postcode{P.O.Box 2713}, \country{Qatar}}}

\affil[2]{\orgdiv{Graduate School of Information Science}, \orgname{University of Hyogo}, \orgaddress{ \city{Kobe 650-0047}, \country{Japan}}}

\abstract{Stroke remains a significant global health concern, necessitating precise and efficient diagnostic tools for timely intervention and improved patient outcomes. The emergence of deep learning methodologies has transformed the landscape of medical image analysis. Recently, \emph{Transformers}, initially designed for natural language processing, have exhibited remarkable capabilities in various computer vision applications, including medical image analysis. This comprehensive review aims to provide an in-depth exploration of the cutting-edge Transformer-based architectures applied in the context of stroke segmentation. It commences with an exploration of stroke pathology, imaging modalities, and the challenges associated with accurate diagnosis and segmentation. Subsequently, the review delves into the fundamental ideas of Transformers, offering detailed insights into their architectural intricacies and the underlying mechanisms that empower them to effectively capture complex spatial information within medical images. The existing literature is systematically categorized and analyzed, discussing various approaches that leverage Transformers for stroke segmentation. A critical assessment is provided, highlighting the strengths and limitations of these methods, including considerations of performance and computational efficiency. Additionally, this review explores potential avenues for future research and development.}

\keywords{stroke segmentation, vision Transformer, deep learning, medical imaging}

\maketitle
\section{Introduction}

Stroke, a cerebrovascular disease, stands as the second leading cause of morbidity and mortality worldwide, impacting over 100 million people globally \cite{feigin2022world}. It transpires when there is an abrupt disruption in the blood supply to the brain, resulting in the damage or death of neuro cells. This occurrence can be attributed to two primary reasons: a blockage in the blood vessels, referred to as ischemic stroke, and the rupture of vessels leading to bleeding into surrounding tissues, known as hemorrhagic stroke \cite{grysiewicz2008epidemiology}. The consequences of stroke on patients can be profound, often resulting in physical disabilities and cognitive impairments \cite{meyer2015systematic, dimyan2011neuroplasticity}. This underscores the importance of accurate and timely diagnosis for effective treatment and improved patient outcomes.

\begin{figure*}
  \centering
  \includegraphics[width=0.75\textwidth]{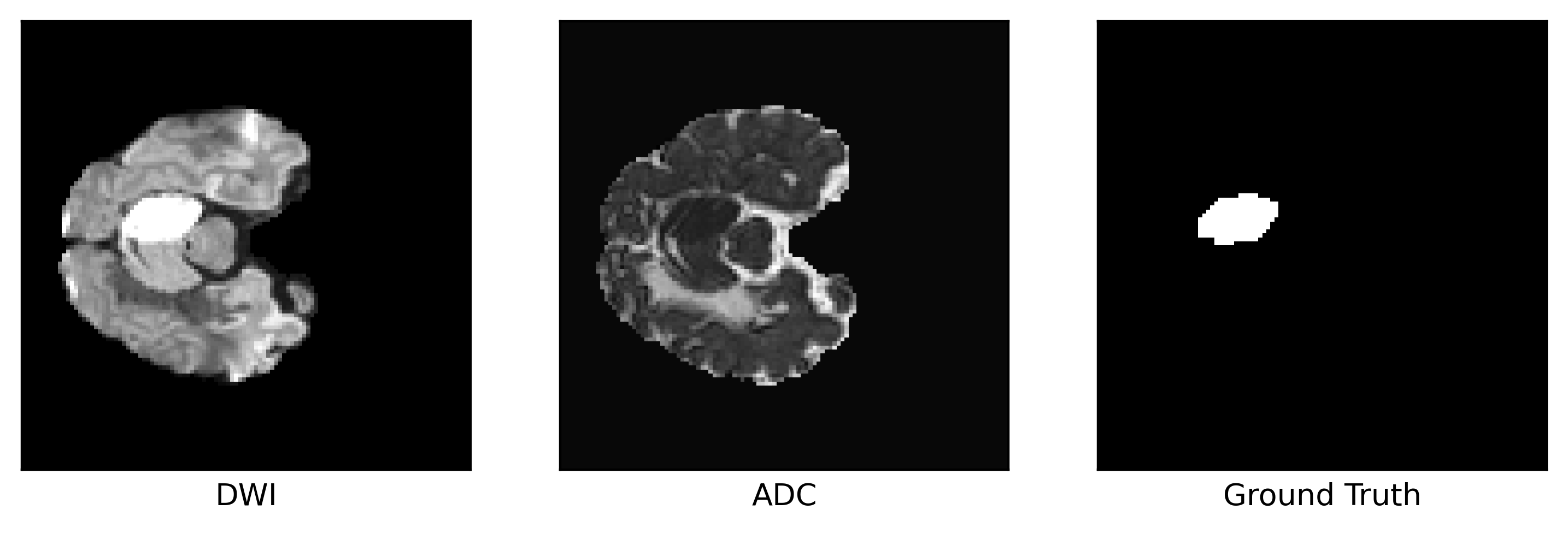}
  \caption{Sample of stroke infarct on Diffusion-weighted and Apparent diffusion coefficient MRI with the annotation, from \cite{hernandez2022isles}. The infarct appears hyperintense in DWI and hypointense in ADC. }
  \label{fig:mri}
\end{figure*}

\begin{figure}
\centering
    \includegraphics[width=0.45\textwidth]{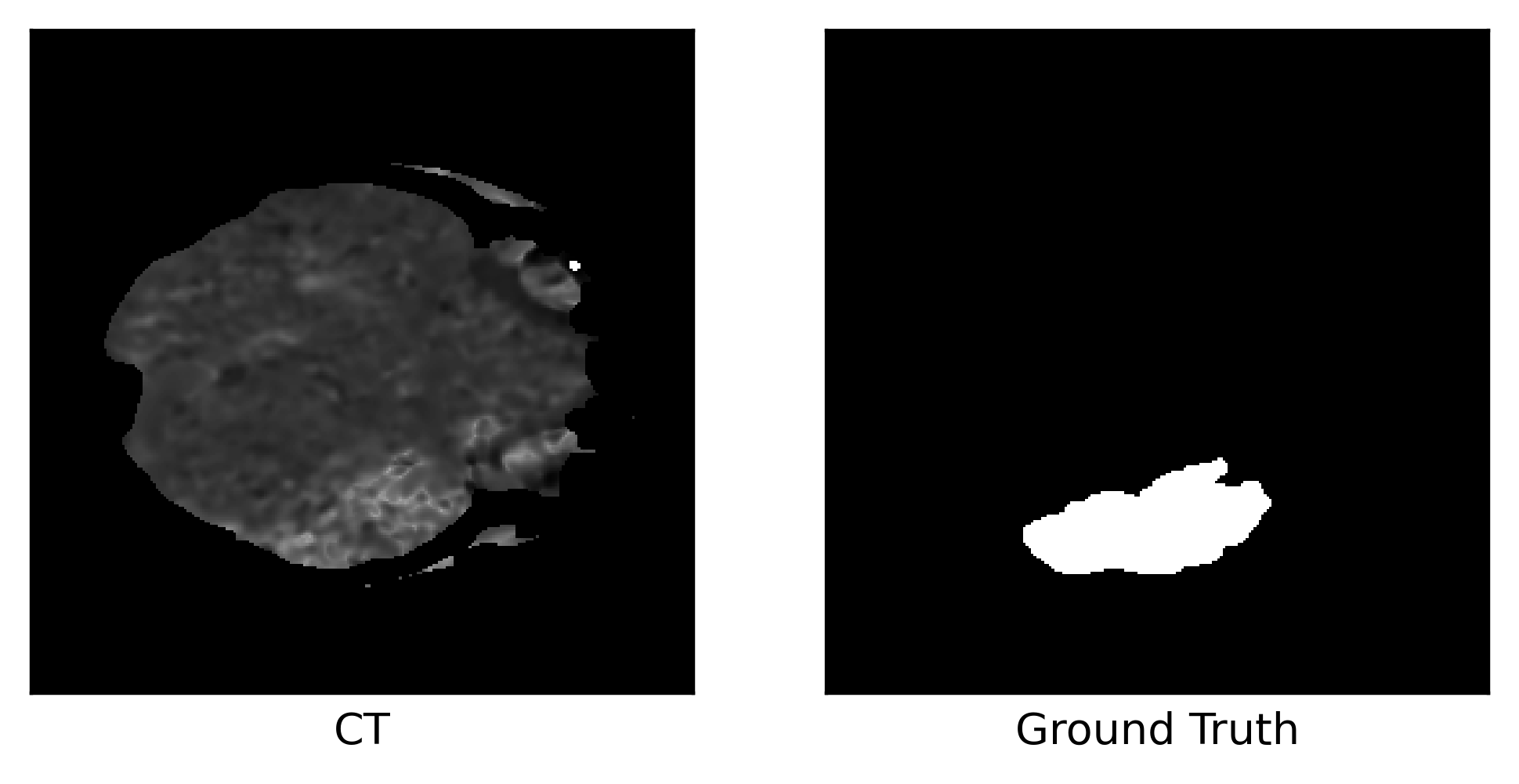}
    \caption{Sample of stroke infarct on CT images with the annotation, from \cite{cereda2016benchmarking, hakim2021predicting}.}
    \label{fig:ct}
\end{figure}

Stroke patients typically undergo neuroimaging techniques to distinguish between ischemic and hemorrhagic strokes. This differentiation can be achieved through magnetic resonance imaging (MRI) and computed tomography (CT), each offering distinctive insight into the condition of the brain \cite{goldstein2005patient}. MRI offers excellent soft tissue contrast for the brain, and when diagnosis is uncertain, it can be more informative than CT \cite{hwang2012comparative, chalela2007magnetic, fiebach2002ct}, providing information on stroke location \cite{flossmann2008reliability}, timing \cite{aoki2010flair}, and mechanism \cite{wessels2006contribution}. Diffusion-weighted imaging (DWI) and perfusion-weighted imaging (PWI) within the MRI protocol offer valuable information on the extent and impact of stroke on brain tissue \cite{simonsen2015sensitivity}. Refer to Figure \ref{fig:mri} for an illustration showing a stroke infarct sample in two distinct magnetic resonance modalities along with the corresponding annotation. Additionally, refer to Figure \ref{fig:ct} for CT images accompanied by corresponding annotations.

Stroke segmentation plays an essential role in the diagnostic process as well as treatment planning by providing spatial information about affected areas of the brain and the extent of damage. Traditional methods of stroke diagnosis, often based on manual interpretation of medical images, prove to be time-consuming and susceptible to human error. The inherent variability in the size, shape, and location of strokes, compounded by artifacts and noise present in the imaging data, presents substantial challenges for automated analysis, rendering it a challenging task. Furthermore, the need for real-time or near-real-time diagnosis in stroke cases demands algorithms that are not only accurate but also computationally efficient. As such, the development of accurate and automatic methods for stroke segmentation remains a prominent focus in the research domain.

The field of medical image analysis has witnessed a transformative evolution with the advent of deep learning techniques \cite{zhou2019handbook}. Deep learning models, with their ability to automatically learn intricate patterns from vast amounts of data, have shown promising results in various medical imaging tasks, including stroke segmentation \cite{zhang2022application}.  Convolutional Neural Networks (CNNs) \cite{o2015introduction}, a class of deep learning models, have demonstrated remarkable success in tasks such as image classification \cite{huang2017densely, hu2018squeeze}, object detection \cite{wang2017fast}, segmentation \cite{chen2017deeplab}, and registration \cite{balakrishnan2019voxelmorph, jia2022u}. These models, with their hierarchical feature learning capabilities, have significantly improved the accuracy and efficiency of medical image analysis. However, the inherent limitations of CNNs in capturing long-range dependencies and contextual information in images have led to the exploration of alternative architectures, including Transformers \cite{li2023transforming}.

Originally proposed for natural language processing tasks \cite{vaswani2017attention}, Transformers have gained widespread attention in the computer vision community. Unlike traditional convolutional approaches, transformers process input data in a parallel and non-sequential manner, allowing them to capture complex spatial relationships and contextual dependencies effectively. The self-attention mechanism in Transformers enables them to weigh different parts of the input data differently, making them particularly suitable for tasks requiring a global understanding of the data, such as medical image analysis.

We conducted a comprehensive search across electronic databases, including PubMed, IEEE Xplore, and Google Scholar. Our search employed diverse queries to compile lists of published works, combining terms like "CNN," "deep learning," "Transformer," and "neural network" with stroke-specific terms such as "ischemic stroke," "stroke segmentation," and "stroke detection." Additionally, we manually explored the reference lists of relevant articles to identify additional studies. We included studies that satisfied the following criteria: (1) publication in English; (2) focus on stroke segmentation; (3) utilization of deep learning techniques, particularly emphasizing Transformer-based architectures; and (4) reporting of quantitative results on the model's performance.

We excluded studies that: (1) were unrelated to stroke segmentation; (2) did not achieve high segmentation performance; and (3) presented papers with identical methodologies where their contributions were negligible. However, in cases where the performance of all proposed pipelines was low for certain datasets, we kept the superior ones. Due to the relatively limited number of Transformer-based networks addressing stroke segmentation in the existing literature, we encompassed all available publications in our study. It is crucial to acknowledge that our review might have unintentionally omitted some noteworthy papers related to CNN-based architectures. Nevertheless, our primary objective was to offer an overview of the contributions in utilizing vision Transformers for stroke segmentation purposes.

\begin{figure}
  \centering
  \includegraphics[width=0.8\textwidth]{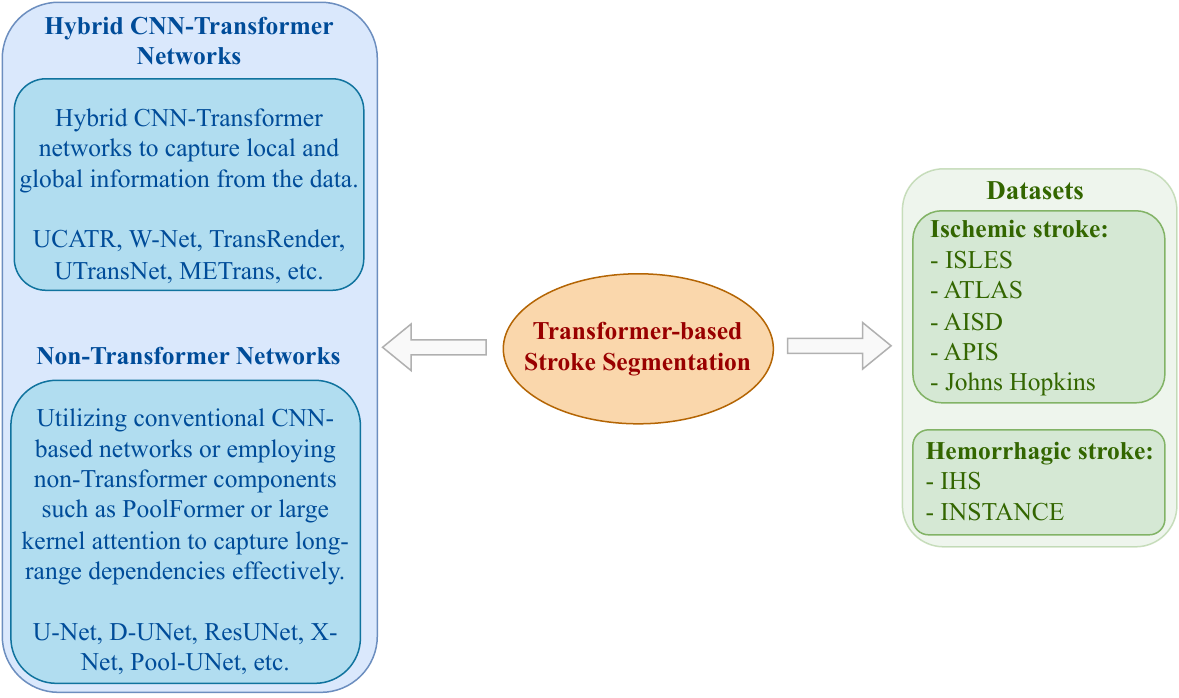}
  \caption{Overview of key aspects covered in this review paper. }
  \label{fig:abstract}
\end{figure}

In this review, we have discussed the applications of Transformers in stroke segmentation, exploring innovative methodologies developed to address the challenges posed by stroke diagnosis. We systematically reviewed the existing literature, analyzing different Transformer-based architectures, their integration with traditional deep learning techniques, and their performance in stroke-related tasks. Through this comprehensive review, we aimed to provide insight into the current state of the art, highlight the strengths and limitations of Transformers in stroke segmentation, and identify potential avenues for future research and development. Figure \ref{fig:abstract} provides a graphical abstract highlighting the key aspects discussed in this paper, encompassing available datasets for stroke segmentation and the diverse deep architectures employed in this context.

\section{Fundamentals of Transformers}

\subsection{Architectural Components of Transformers}

Transformers, introduced in the context of natural language processing, consist of fundamental architectural components that distinguish them from traditional CNN-based models. The core elements of Transformers include self-attention mechanisms and position-wise feed-forward networks. Self-attention enables the model to weigh input elements differently, capturing contextual dependencies irrespective of their positions in the sequence. This mechanism allows Transformers to model long-range dependencies efficiently, making them well-suited for tasks requiring a global understanding of the data. Position-wise feed-forward networks introduce non-linear transformations, enhancing the model's capacity to learn complex patterns within the data.
\subsubsection{Self-Attention}
The self-attention (SA) mechanism used in Transformers is a crucial component that empowers the model to capture the long-range dependencies between various parts of the input data. This is accomplished through a process in which each element (token) in the input sequence attends to every other element, calculating its representation based on the information from all other elements.

To compute the self-attention, the input sequence $X\in\mathbb{R}^{N{\times}C}$ is projected into a query $Q\in\mathbb{R}^{N{\times}D}$, a key $K\in\mathbb{R}^{N{\times}D}$, and a value $V\in\mathbb{R}^{N{\times}D_v}$ using three trainable projection layers $W^Q\in\mathbb{R}^{C{\times}D}, W^K\in\mathbb{R}^{C{\times}D}, W^V\in\mathbb{R}^{C{\times}D_v}$, respectively. Then, the corresponding attention matrix $A\in\mathbb{R}^{N{\times}N}$, which represents the affinity of the query and the key, can be calculated by:
 \begin{equation}
     A(Q, K) = Softmax(\frac{Q{\times}K^T}{\sqrt{D}})
 \end{equation}
The attention matrix connects all elements, which allows the handling of long-range dependencies. Subsequently, the calculated attention matrix is applied to the value $V$, resulting in the output $Z\in\mathbb{R}^{N{\times}D_V}$:
\begin{equation}
    Z = SA(Q, K, V) = A(Q, K){\times}V
\end{equation}

\subsubsection{Multi-Head Self-Attention}
In multi-head self-attention (MSA), multiple SA blocks (heads) are performed in parallel to produce multiple output maps. The final output is a concatenation and projection of all outputs of SA blocks. This enables better modeling of complex dependencies between different elements in the input. For $H$ number of heads, each head has its learnable weight matrices, $ \{W^{(Q_i)}, W^{(K_i)}, W^{(V_i)}; i=1,..., H\}$.
\begin{equation}
\begin{split}
        Z_i = SA(Q_i, K_i, V_i) = Softmax(\frac{Q_i{\times}K_i^T}{\sqrt{D/H}}), \\
    MSA(Q, K, V) = Concat (Z_1, Z_2,..., Z_H) W^0
\end{split}
\end{equation}

where $W^0$ is a linear projection that aggregates the outputs of all attention heads. It is noteworthy that a larger number of heads does not necessarily lead to better performance \cite{dosovitskiy2020image}.

\subsection{Vision Transformer Pipeline}
The Vision Transformer \cite{dosovitskiy2020image}, or ViT, is a Transformer-like architecture introduced for image classification tasks. The main paradigm in the ViT is that tokens are created from the flattened patches of the image. Let $X$ be a 3D image volume ($X\in\mathbb{R}^{(H{\times}W{\times}L{\times}C)}$), where $(H, W, L)$ represents the image dimensions, and $C$ is the number of channels. The image is divided into $N$ patches, which can overlap or not overlap, with each patch having a size of $(P, P, P)$. Then, a sequence is created from the flattened form of these patches $x_P\in\mathbb{R}^{N{\times}P^3C}$ and projected into a $D$ dimensional space $\hat{x}$. To preserve positional information, a positional embedding was added, resulting in the input of the Transformer encoder, denoted as $x$:
\begin{equation}
    x = \hat{x} + E_{pos}, E_{pos}\in\mathbb{R}^{N{\times}D}
\end{equation}

\begin{figure}
    \centering
    \includegraphics[width=.6\textwidth]{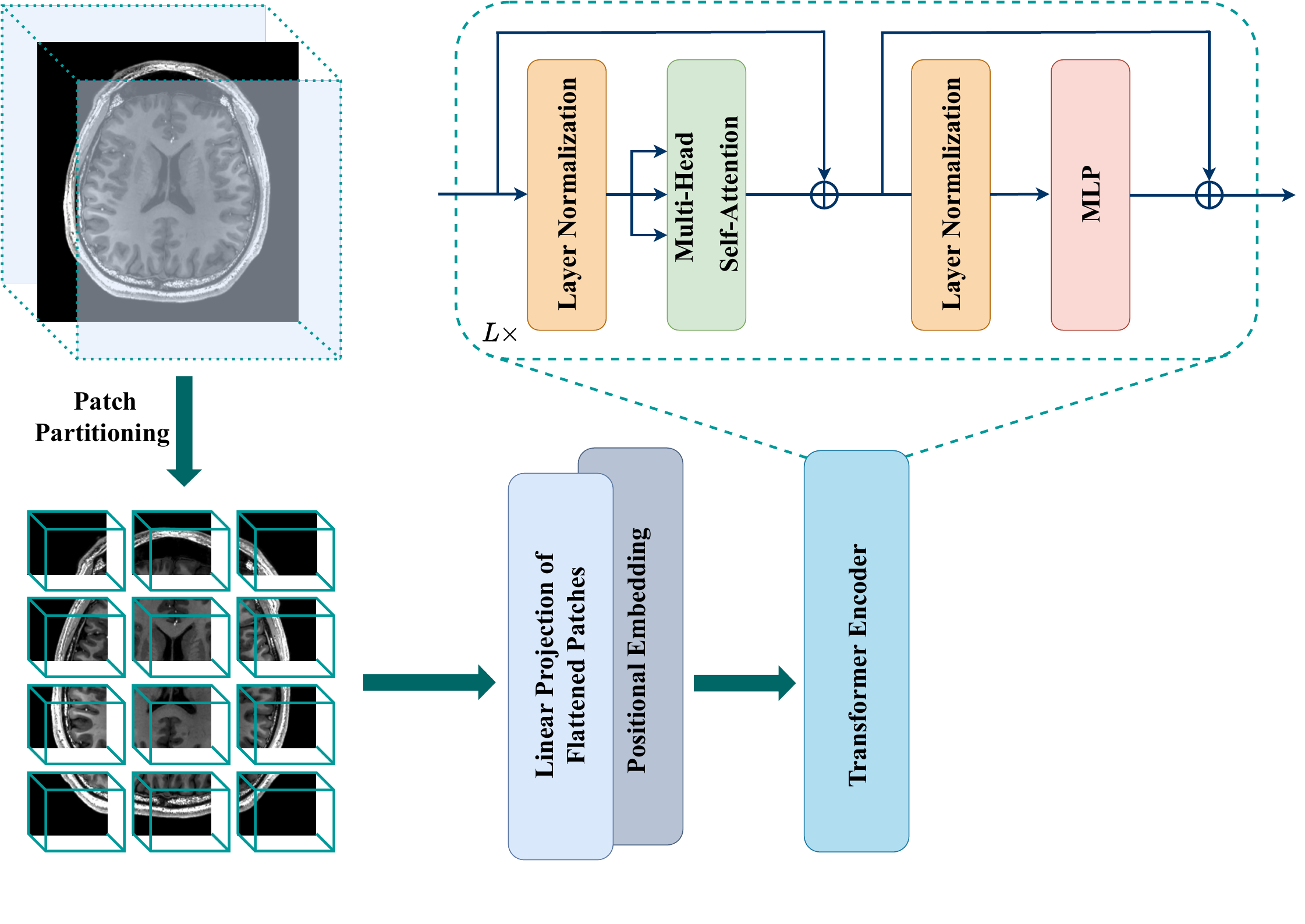}
    \caption{Overview of Vision Transformer (ViT) and the Transformer encoder.}
    \label{fig:vit}
\end{figure}

The subsequent tokens were inputted into a Transformer encoder comprising $L$ stacked base blocks. Each base block included multi-head self-attention and a multilayer perceptron (MLP) with layer normalization (LN), and residual connections were employed following each block. A depiction of ViT and the Transformer encoder is shown in Fig. \ref{fig:vit}

\begin{figure*}
    \centering
    \includegraphics[width=.9\textwidth]{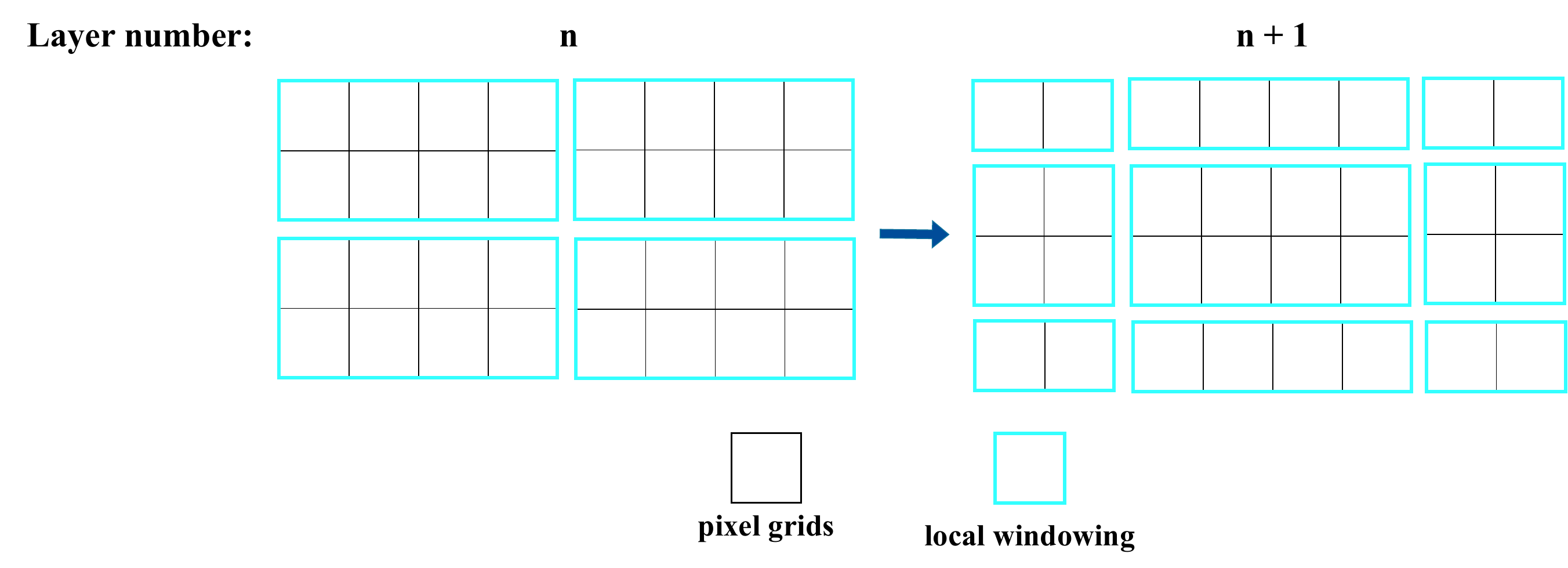}
    \caption{ The cyclic shift of the local window for Shifted Windows-based self-attention computation. The self-attention is computed in each local window.}
    \label{fig:swin}
\end{figure*}

It is noteworthy that the computational complexity of calculating the Softmax within the MSA blocks grows quadratically as the input sequence length increases \cite{dosovitskiy2020image}. This limitation could restrict its practical use, especially when dealing with high-resolution medical images. The introduction of the "Shifted Windows" idea in the Swin Transformer \cite{liu2021swin} improved the efficiency of MSA calculations. Unlike ViT \cite{dosovitskiy2020image}, which computes the relationship between one token and all others in every step of the self-attention calculations, Swin Transformer restricts self-attention calculations to non-overlapping local windows. It also enables cross-window connections and maintains linear computational complexity relative to the image size. Refer to Figure \ref{fig:swin} for a visual representation of how the Shifted Windows concept partitions an input feature map with dimensions of $4{\times}8$ pixels, using a window size of $2{\times}4$. Additionally, it utilizes a hierarchical structure and generates feature maps at multiple resolutions through the incorporation of patch-merging layers.

\subsection{Adaptations for Medical Image Analysis}
Adapting Transformers for medical image analysis involves several considerations. One significant challenge is the high dimensionality of medical images, often requiring substantial computational resources for processing, especially for 3D images. Researchers have explored techniques such as patch-based processing \cite{liu2021swin} and efficient attention mechanisms \cite{xiong2021nystromformer,rao2021dynamicvit} to mitigate this challenge. Another significant challenge lies in ensuring the generalizability of the proposed network. The pipeline must demonstrate robustness when tested on unseen data acquired from different imaging scanners or centers. The inherent variability among images obtained from different vendors, even when imaging the same subject, can result in a noticeable reduction in performance accuracy. Addressing and managing this variability is essential for handling the challenges associated with generalization. Various applications require specific conditions to be satisfied, and these conditions can vary significantly between different applications. The design of networks, especially Transformer-based architectures, should be approached with careful consideration based on the unique nature and requirements of each application.

Several architectures have been developed utilizing exclusively Transformer models \cite{cao2022swinunet, karimi2021convolution-free, wang2021Ted-net}. DAE-Former \cite{azad2023dae} is a dedicated Transformer architecture proposed for medical image segmentation, featuring dual attention blocks in both the encoder and decoder, along with cross-attention blocks in the skip connections to optimize segmentation results. The key elements of dual attention blocks include efficient attention \cite{shen2021efficient}, employed to reduce computational complexity from quadratic to linear. Additionally, transpose attention \cite{ali2021xcit} is incorporated into these blocks to capture channel attention. This architectural choice is based on empirical evidence suggesting that combining spatial and channel attention enhances the model's capacity to capture more contextual features \cite{guo2022attention}.

Pure Transformer architectures demonstrate certain limitations when compared with hybrid architectures that effectively capture both local and global information. Combining Transformers with CNNs in hybrid architectures leverages the strengths of both models, allowing Transformers to capture global context and CNNs to learn local features \cite{chen2021transunet}. These adaptations have paved the way for the successful application of Transformers in tasks like stroke segmentation, addressing the unique requirements of medical image analysis. Hybrid Transformer-CNN models offer flexibility in the placement of the Transformer component.

In Swin UNETR model \cite{hatamizadeh2021swinunetr}, the Swin Transformer took on the role of the encoder, and the encoded features from the Transformer were combined with the CNN-based decoder at various levels and resolutions. UNETR \cite{hatamizadeh2022unetr} consisted of a Transformer-based encoder and a CNN-based decoder, featuring skip connections composed of convolutional-based blocks. In TransUNET model \cite{chen2021transunet}, the Transformer was integrated into the encoder, where it processed tokenized image patches derived from a CNN-generated feature map. TransFuse \cite{zhang2021transfuse} introduced a unique approach, employing a dual-branch encoder, one branch based on CNN and the other solely on the Transformer. A novel technique, called BiFusion, fused multi-level features extracted from both branches. In nnFormer \cite{zhou2021nnformer}, a combination of interleaved convolution and self-attention operations was employed. Additionally, nnFormer utilized skip attention, akin to the traditional concatenation approach seen in skip connections within UNet-like architectures. Transformers have proven their effectiveness when utilized as the upsampling components within the decoder section \cite{li2022more}.

\begin{figure*}
  \centering
  \includegraphics[width=1\textwidth]{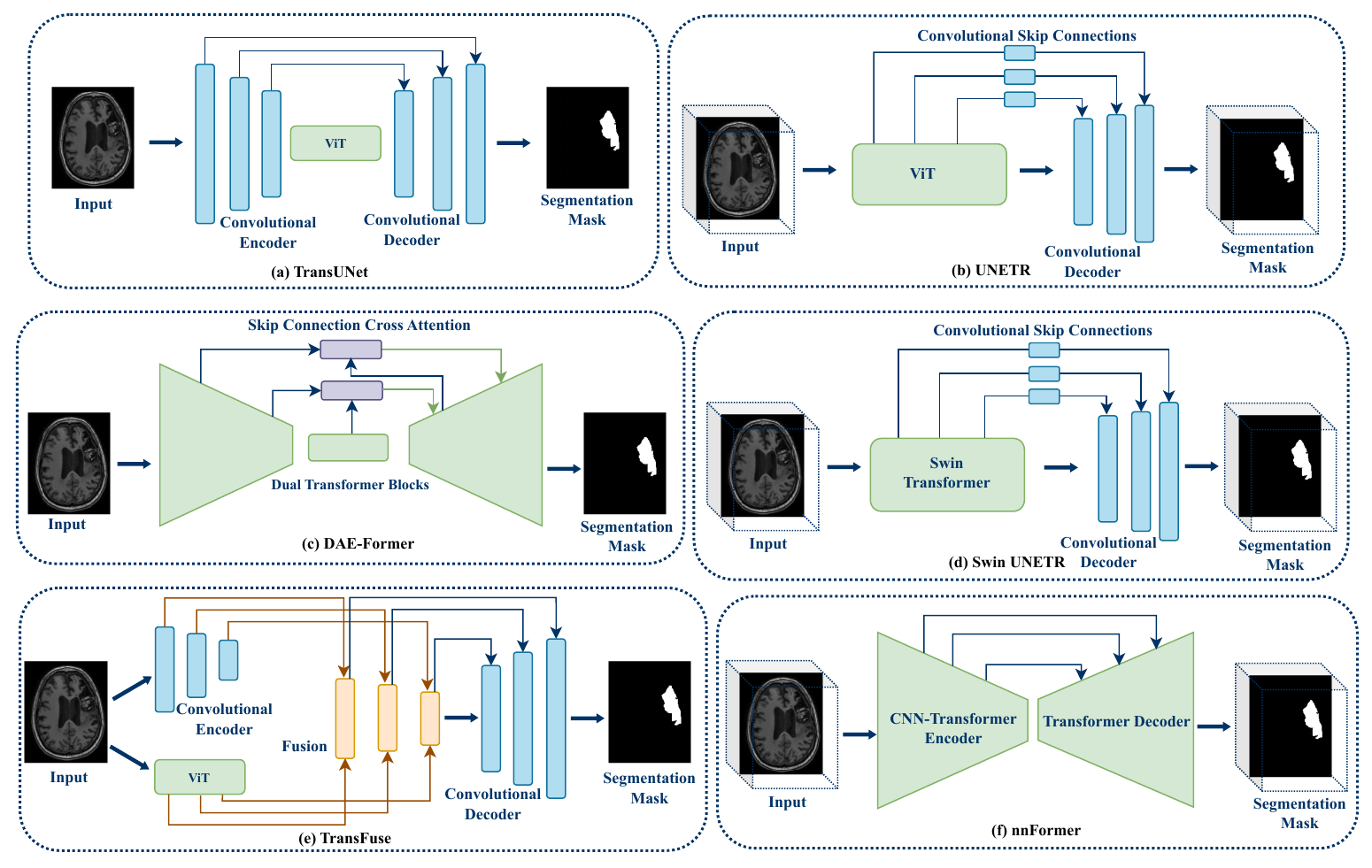}
  \caption{Transformer-based networks as segmentation model architectures. The schematic indication of whether the network is designed for a 2D or 3D image is presented in each figure. (a) TransUNet \cite{chen2021transunet} incorporates Transformer blocks as additional encoders to model bottleneck features. (b)  UNETR \cite{hatamizadeh2022unetr} employs the Transformer as the primary encoder path, combining it with a CNN decoder and skip connections to create a hybrid model. (c) DAE-Former \cite{azad2023dae} integrates Transformers into dual attention blocks in both the encoder and decoder, enhancing the architecture with cross-attention in skip connections. (d) Swin UNETR \cite{hatamizadeh2021swinunetr} features the Swin Transformer in the encoder section, complemented by a CNN-based decoder and skip connections to form a hybrid model. (e) TransFuse \cite{zhang2021transfuse} fuses the Transformer and CNN encoder to connect the decoder. (f) nnFormer \cite{zhou2021nnformer} incorporates Transformers in both the encoder and decoder for its architecture. }
  \label{fig:architectures}
\end{figure*}

The Fully Convolutional Transformer (FCT) \cite{tragakis2023fully} was conceived to harness the strengths of CNNs for local feature representation and capitalize on Transformers' proficiency in capturing long-range dependencies. The utilization of depth-wise convolutions in the projection layer obviates the necessity for positional embedding addition. Following the extraction of overlapping patches from an image, patch-based embeddings are incorporated, and MSA is subsequently calculated on these patches. In the FCT framework, a multi-branch convolutional paradigm is embraced to enhance spatial context. In this context, one layer applies a spatial convolution to the MSA output with a small kernel size, while other layers employ dilated convolutions with larger receptive fields. The integration of these outputs is facilitated by a fusion module known as Wide-Focus. Figure \ref{fig:architectures} illustrates several typical Transformer-based architectures for medical image segmentation, which have served as inspiration for many related models.

\section{Datasets}
In this section, we introduced available datasets for stroke infarct segmentation, encompassing both ischemic and hemorrhagic strokes across both CT and MRI modalities. Each dataset comprises various modalities with a different number of cases. Table \ref{tab:data} provides a summary of these datasets.
\subsection{ISLES Dataset}
The Ischemic Stroke Lesion Segmentation (ISLES) challenge offers publicly available datasets, released in 2015, 2017, 2018, and 2022. The objectives of the challenge varied each year, and distinct image modalities were provided in each one.
   \subsubsection{ISLES 2015 }
ISLES 2015 \cite{maier2017isles2015} is a publicly available dataset comprising two distinct sub-challenges: Sub-Acute Ischemic Stroke Lesion Segmentation (SISS) and Stroke Perfusion Estimation (SPES).

ISLES 2015-SISS consisted of 64 sub-acute ischemic cases, with 28 cases allocated for training and 36 cases for testing with a voxel size of $1 {mm}^3$. These cases were collected from two different medical centers with variations in image resolution. Each case within the dataset was accompanied by four MRI modalities, namely T1-weighted, T2-weighted, DWI, and Fluid attenuated inversion recovery (FLAIR). Preprocessing steps, including skull-stripping and resampling to an isotropic space, were applied, and all modalities were registered to the FLAIR modality as a reference. Both the training and testing datasets included instances of single and multi-focal lesions, as well as large and small lesions.

ISLES 2015-SPES comprised 50 acute ischemic cases, with 30 cases designated for training and 20 cases for testing, with a voxel size of $2 {mm}^3$. Each case was accompanied by seven distinct image modalities, namely T1 contrast-enhanced (T1c), T2, DWI, cerebral blood flow (CBF), cerebral blood volume (CBV), time-to-peak (TTP), and time-to-max (Tmax). To ensure consistency, all modalities were registered on the T1c modality.

\subsubsection{ISLES 2017}
This dataset \cite{winzeck2018isles2017} represents an extension of the ISLES 2016 stroke lesion segmentation challenge, notable for its expansion in the number of acute ischemic cases. Specifically, the dataset increased from 35 training and 19 testing cases to 43 training and 32 testing cases. Furthermore, this dataset introduced a new set of MRI modalities, including ADC, rBF, rBV, MTT, Tmax, TTP, and raw PWI, distinguishing it from the previously utilized modalities in the 2015 version. The preprocessing steps in ISLES 2017 were more concise, primarily encompassing registration and skull-stripping, while the voxel size and image resolutions exhibited variations.

\subsubsection{ISLES 2018 }
The primary objective of the ISLES 2018 dataset \cite{cereda2016benchmarking, hakim2021predicting} was to perform the segmentation of stroke lesions using computed tomography perfusion (CTP) images, guided by annotations derived from DWI images, which are considered the standard image modalities. The dataset encompasses information from 103 acute ischemic cases with MRI images acquired within a 3-hour window of CTP. For training, 63 cases were designated, while the remaining 40 were reserved for testing. The input data for the algorithms consisted of various perfusion maps, including cerebral blood volume (CBV), cerebral blood flow (CBF), MTT, and Tmax.

\subsubsection{ISLES 2022}
The dataset \cite{hernandez2022isles}, sourced from three distinct medical centers, comprised information from 400 acute and sub-acute ischemic cases. Notably, this dataset featured a diverse range of infarct patterns, and a high degree of variability in terms of lesion size and location, with a mean number of 9.289 and a maximum of 126 unconnected ischemic regions per scan. The dataset exhibited heterogeneity due to the utilization of three different imaging devices, which serves as a valuable criterion for assessing the generalization of proposed methods. Modalities included in this dataset encompass DWI, ADC, and FLAIR images.

\subsection{ATLAS Dataset}
The Anatomical Tracings of Lesions After Stroke (ATLAS) v2.0 dataset \cite{liew2022atlas} contained high-resolution T1-weighted images for the segmentation of acute, sub-acute, and chronic stroke lesions. Significantly, this dataset was more extensive, containing more than four times the data volume of its predecessor, ATLAS v1.2. Aggregated from multiple centers worldwide, the dataset included data from 1,271 cases. Of these, 655 cases were allocated for training, 300 cases offered images only with hidden segmentation masks, and an additional 316 cases were entirely withheld to assess the generalizability of the proposed methods. The preprocessing steps applied to this dataset involved intensity normalization and registration on the MNI-152 template.

\subsection{AISD Dataset}
The Acute ischemic stroke dataset (AISD) \cite{liang2021symmetry} comprised paired CT-MRI data for 397 acute ischemic stroke cases. The dataset included Non-Contrast-enhanced CT (NCCT) scans and DWI scans, which were acquired within 24 hours of the CT images. The segmentation labels were derived from the MRI scans, which served as the standard for this purpose.

\subsection{APIS Dataset}
The APIS dataset \cite{gomez2023apis} was designed as a paired CT-MRI dataset with the objective of ischemic stroke lesion segmentation, utilizing NCCT images and annotations from ADC scans. The training set comprised 60 pairs of CT-MRI data, while the testing phase involved 36 NCCT scans exclusively. All cases underwent preprocessing steps, including skull-stripping and registration onto the ADC scans, to ensure data consistency and alignment.

\subsection{Johns Hopkins University's Dataset }
This dataset comprised 2888 MRI datasets from cases involving acute and early subacute stroke patients, along with corresponding annotations \cite{liu2023large}. The data were collected over 10 years using 11 MRI scanners. For all patients, DWI images, B0, and ADC were provided, and nearly 98.8\% of patients had additional MRI modalities, including T1, high-resolution T1 MPRAGE, T2, FLAIR, SWI, and PWI. DWI images were registered onto the standard MNI space, subjected to skull-stripping, and resampled to a voxel size of $1 {mm}^3$. The considerable diversity within this dataset renders it an excellent benchmark for evaluating proposed methods in the context of stroke segmentation. Nevertheless, it is important to note that access to this dataset is subject to certain restrictions.

\subsection{Intracranial Hemorrhage Segmentation (IHS) Dataset}
The dataset comprised non-contrast CT scans from 36 patients diagnosed with various types of intracranial hemorrhage, including intraventricular, intraparenchymal, subarachnoid, epidural, and subdural hemorrhage \cite{hssayeni2020intracranial}. Each scan was characterized by an average of 30 slices with a thickness of $5 {mm}$. Annotations for the dataset were provided by two radiologists. Collected in 2018, this dataset is accessible from PhysioNet \cite{goldberger2000physiobank} with certain restrictions.

\subsection{INSTANCE Dataset}
For the Intracranial Hemorrhage Segmentation on Non-Contrast Head CT (INSTANCE) challenge \cite{li2023state}, a dataset comprising non-contrast CT scans from 200 patients diagnosed with various types of intracranial hemorrhage was assembled. The dataset allocation for different phases was as follows: 100 scans were designated for the training phase, 30 cases without ground truth labels were set aside for validation, and the remaining 70 cases were reserved for the final evaluation. The image size for each slice was $512 {\times} 512$, with the number of slices varying from 20 to 70 for each case. While the pixel size in each slice was $0.42 {mm}^2$, providing a good inter-slice resolution, the slice thickness was $5 {mm}$, resulting in a lower inter-slice resolution.

 \begin{table*}
     \centering
     \caption{Summary of available datasets information for stroke segmentation}
     \resizebox{0.8\textwidth}{!}{
     \begin{tabular}{|l|l|c|l|}
     \hline
    Dataset & Disease &  Number of Cases & Target Modality\\
         \hline
         ISLES 2015-SPES & Acute ischemic stroke & 50 & MRI\\ [0.03cm]
      
         ISLES 2015-SISS & Subacute ischemic stroke & 64 & MRI\\ [0.03cm]
         
         ISLES 2017 & Acute ischemic stroke & 75 & MRI\\ [0.03cm]
          
         ISLES 2022 & Acute and subacute ischemic stroke & 400 & MRI\\ [0.03cm]
         
         ATLAS v2.0 & Acute, subacute, and chronic ischemic stroke & 1271 & MRI\\ [0.03cm] 
         
         Johns Hopkins University & Acute and early subacute ischemic stroke & 2888 & MRI \\ [0.03cm] 
         
         ISLES 2018 & Acute ischemic stroke & 103 & CT\\ [0.03cm]
         
         AISD & Acute ischemic stroke & 397 & CT \\ [0.03cm] 
         
         APIS & Ischemic stroke & 96 & CT \\ [0.03cm]
         
         IHS & Hemorrhagic stroke & 36 & CT \\ [0.03cm]
         
         INSTANCE & Hemorrhagic stroke & 200 & CT \\ [0.03cm] 

         \hline
     \end{tabular}
     }
     \label{tab:data}
 \end{table*}

\section{Performance Evaluation for Stroke Segmentation}
Quantitative analysis of a segmentation process, evaluating its effectiveness in categorizing pixels or voxels into desired classes, is a crucial component of model evaluation. Many commonly used metrics rely on pixel-wise or voxel-wise calculations. The simplest method for assessing performance is through overall accuracy, defined as:
\begin{equation}
    Overall Accuracy = \frac{TP + TN}{TP + TN + FP + FN}
\end{equation}

where, TP, TN, FP, and FN represent true positives, true negatives, false positives, and false negatives, respectively. However, overall accuracy may not provide sufficient insights, especially in imbalanced tasks such as stroke segmentation. To address this limitation, alternative metrics are widely used for a more nuanced evaluation of imbalanced semantic segmentation performance. One prominent metric is the Dice Similarity Coefficient (DSC), which measures the overlap between the predicted segmentation and the ground truth, ranging from 0 (no overlap) to 1 (perfect overlap). It is calculated as:
\begin{equation}
    DSC= \frac{2{\times}TP}{2{\times}TP + FP + FN}
\end{equation}

Another valuable metric is the Intersection over Union (IoU), also known as the Jaccard Index, which assesses the ratio of the overlap to the total combined area, ranging from 0 (no similarity) to 1 (perfect similarity). The IoU is computed as:
\begin{equation}
    IoU= \frac{TP}{TP + FP + FN}
\end{equation}

Both DSC and IoU provide comprehensive insights into the agreement between the predicted segmentation and the ground truth, making them particularly useful in scenarios with an imbalanced class distribution. There are also other commonly used metrics to measure stroke segmentation performance, including Precision, which assesses the accuracy of the positive predictions, Recall/Sensitivity, which gauges the ability to capture positive instances, and F1-score, a metric that strikes a balance between precision and recall by calculating their harmonic mean.
\begin{equation}
    Precision = \frac{TP}{TP + FP}
\end{equation}

\begin{equation}
    Recall = \frac{TP}{TP + FN}
\end{equation}

\begin{equation}
    F1-score = \frac{2{\times}Recall{\times}Precision}{Recall + Precision}
\end{equation}

F1-score is a specific form of the general $F_{\beta}-score$, in which the parameter $\beta$ controls the trade-off between recall and precision, particularly useful when there is uneven importance assigned to precision and recall. The formula for the $F_{\beta}-score$ is as follows:
\begin{equation}
  F_{\beta}-score = \frac{(1 + {\beta}^2){\times}Recall{\times}Precision}{Recall + {\beta}^2 {\times}Precision}  
\end{equation}

Another metric worth considering is the Hausdorff Distance (HD), which measures the maximum distance between two segmentation sets. Utilizing the HD metric reflects the level of dissimilarity between predicted and ground truth boundaries. The HD is computed as follows:
\begin{equation}
    \text{HD}(A, B) = \max\left(\sup_{a \in A} \inf_{b \in B} d(a, b), \sup_{b \in B} \inf_{a \in A} d(a, b)\right)
\end{equation}

Here, $A$ and $B$ represent two sets, and $d(a, b)$ is the distance function between points $a$ in set $A$ and $b$ in set $B$. The formula calculates the HD by determining the maximum of the infimum of distances from points in set $A$ to the nearest point in set $B$ and vice versa. This measurement can be represented in mm or voxel/pixel-based units. Some papers also utilized additional measurements, including Simple Lesion Count (SLC), Volume Difference (VD), Average Volume Difference (AVD), Volumetric Overlap Error (VOE), Relative Volume Difference (RVD), and average symmetric surface distance (ASSD).

\section{Stroke Segmentation using Transformers}
   \subsection{Earlier Approaches for Stroke Segmentation}
The majority of prior studies of stroke lesion segmentation have primarily focused on CNN models, with an emphasis on U-Net-based architectures \cite{clerigues2019acute, basak2020f, khezrpour2022automatic, liu2020attention, kadry2021u}. Notably, the U-Net \cite{ronneberger2015unet} architecture, specifically designed for biomedical image segmentation, exhibits a distinctive U-shaped configuration, comprising an encoder segment dedicated to contextual feature extraction and a decoder segment tailored for accurate localization. The incorporation of skip connections within this architecture facilitated the seamless integration of high-level feature maps derived from the encoder path with fine-grained details from the decoder path, enhancing its segmentation performance. Some research studies have drawn inspiration from the DenseNet architecture for their utilized neural network design \cite{zhang2018automatic}. Table \ref{Tab:cnn} summarizes the performance of selected CNN-based methods for stroke segmentation across various datasets. The inclusion of papers in the table is based on their demonstrated superior performance.

The bilateral quasi-symmetry property of the brain has been utilized in some research studies \cite{liang2021symmetry, wang2016deep, clerigues2020acute, vupputuri2018symmetry}. In \cite{clerigues2020acute} a patch-based deep learning pipeline was proposed, wherein the extraction of patches was carried out with a significant degree of overlap. These patches were subsequently fed into the neural network using a well-balanced sampling strategy, to mitigate issues associated with class imbalance. In \cite{praveen2018ischemic} a stacked sparse autoencoder network for unsupervised feature learning was employed, which was subsequently coupled with a Support Vector Machine (SVM) classifier to classify patches into normal and lesion categories.

The D-UNet \cite{zhou2019d} model was introduced for chronic stroke segmentation, and it incorporated a fusion of 2D and 3D convolutions in the encoder stage via a dimension transform block. This combination of 2D and 3D information facilitated more effective lesion identification. Additionally, a novel loss function called Enhance Mixing Loss (EML) was employed, which is a composite of the Focal loss \cite{lin2017focal} and Dice coefficient loss. By evaluating their proposed method on the ATLAS dataset, they achieved a mean Dice coefficient of 0.5349 for pixel-wise calculation and 0.7231 for voxel-wise calculation, respectively. In \cite{kumar2020csnet} a hybrid Classifier-Segmenter network (CSNet) was introduced. Initially, the images were input into a classifier that distinguished slices with lesions. The chosen slices were then fed into a segmenter network, which utilized a fractal U-Net model for segmentation, and a final voting mechanism was employed to improve segmentation performance.

In \cite{abulnaga2019ischemic} a CNN-based model was introduced that incorporated the pyramid pooling module, as introduced in PSPNet \cite{zhao2017pyramid}. This module was utilized to extract global and local contextual information, enhancing the accuracy of stroke lesion segmentation. It achieves this by capturing global information through the use of varying kernel sizes and aggregating multi-scale region-based context. The best results were achieved by incorporating pretraining and utilizing the Focal Loss on the ISLES 2018 dataset.

In X-Net \cite{qi2019x}, a Feature Similarity Module (FSM) was implemented to capture long-range dependencies, thereby enhancing the segmentation process. This module was employed at the bottleneck between the encoder and decoder to investigate dense long-range contextual information. To reduce network size and control the number of parameters, mitigating the risk of overfitting, depthwise convolutions were also integrated. Their proposed network demonstrated superior performance compared to other architectures, including U-Net, SegNet \cite{badrinarayanan2017segnet}, PSPNet \cite{zhao2017pyramid}, ResUNet \cite{zhang2018road}, 2D Dense-UNet \cite{li2018h}, and DeepLabv3+ \cite{chen2018encoder}, as evaluated on the ATLAS dataset.

In \cite{liu2019efficient} a CNN-based pipeline was proposed that divides the network into two subnetworks. This approach involved using multi-kernels of various sizes to extract feature maps across different receptive fields. Post-processing techniques were also applied to retain edge details in the images and reduce noise. The optimal performance for their proposed pipeline was achieved by employing a dropout rate of 0.1. \cite{bal2023robust} followed a similar approach and incorporated a local pathway and a global pathway within their proposed model, with larger kernel sizes employed in the global pathway to expand the receptive field for extracting long-range dependencies and global information. The best results in their proposed pipeline were obtained through the inclusion of preprocessing and data augmentation for the ISLES 2015 dataset.

\cite{zhang2020ischemic} devised a pipeline centered around a Detection and Segmentation Network (DSN). They utilized a triple-branch architecture to extract predictions for slices in the axial, sagittal, and coronal planes separately. Subsequently, the predicted labels from different slices within each plane were fine-tuned and passed through a fusion module to obtain the final segmentation label. Their approach outperformed architectures such as U-Net, V-Net \cite{milletari2016v}, and DeepMedic \cite{kamnitsas2016deepmedic} in validation using the ISLES-SSIS dataset.

In \cite{huo2022mapping} a model within the nnU-Net framework \cite{isensee2021nnu} was introduced. Their approach incorporated four schemes: a generic U-Net, utilizing a TopK10 loss to improve performance in small lesion segmentation, a residual U-Net, and a self-training U-Net to enhance model diversity. An ensemble method was employed to combine the predicted results from these four networks, followed by post-processing techniques to enhance the results.

In \cite{abramova2021hemorrhagic} a 3D U-Net-based network for hemorrhagic stroke segmentation in CT scans was utilized. To enhance the representation of informative features, they incorporated Squeeze and Excitation (SE) blocks \cite{hu2018squeeze} in the bottleneck and last layers of their network. Through symmetric data augmentation and the implementation of a restrictive patch sampling approach, their proposed architecture achieved a mean Dice coefficient of 0.86 on a clinical dataset consisting of 76 cases.

Pool-UNet \cite{liu2022pool} incorporated SE blocks in a novel module called DSE-ResNet placed in the bottleneck. This module captures interdependencies between channels to provide the most informative features for the decoder. Additionally, they combined the Poolformer structure \cite{yu2022metaformer}, a transformer-like structure utilizing pooling operations, with CNNs to capture both local and global information. Evaluations on the ISLES 2018 dataset demonstrated the superior performance of their proposed architecture compared to architectures such as U-Net, R2UNet \cite{alom2018recurrent}, and TransUNet. \cite{chalcroft2023large} employed Large Kernel Attention \cite{guo2023visual} to capture long-range dependencies, capitalizing on the inherent biases of convolutions. The Large Kernel Attention mechanism comprises a sequence of depth-wise convolutions, dilated depth-wise convolutions, and pointwise convolutions.

\begin{figure}
    \centering
    \includegraphics[width=.65\textwidth]{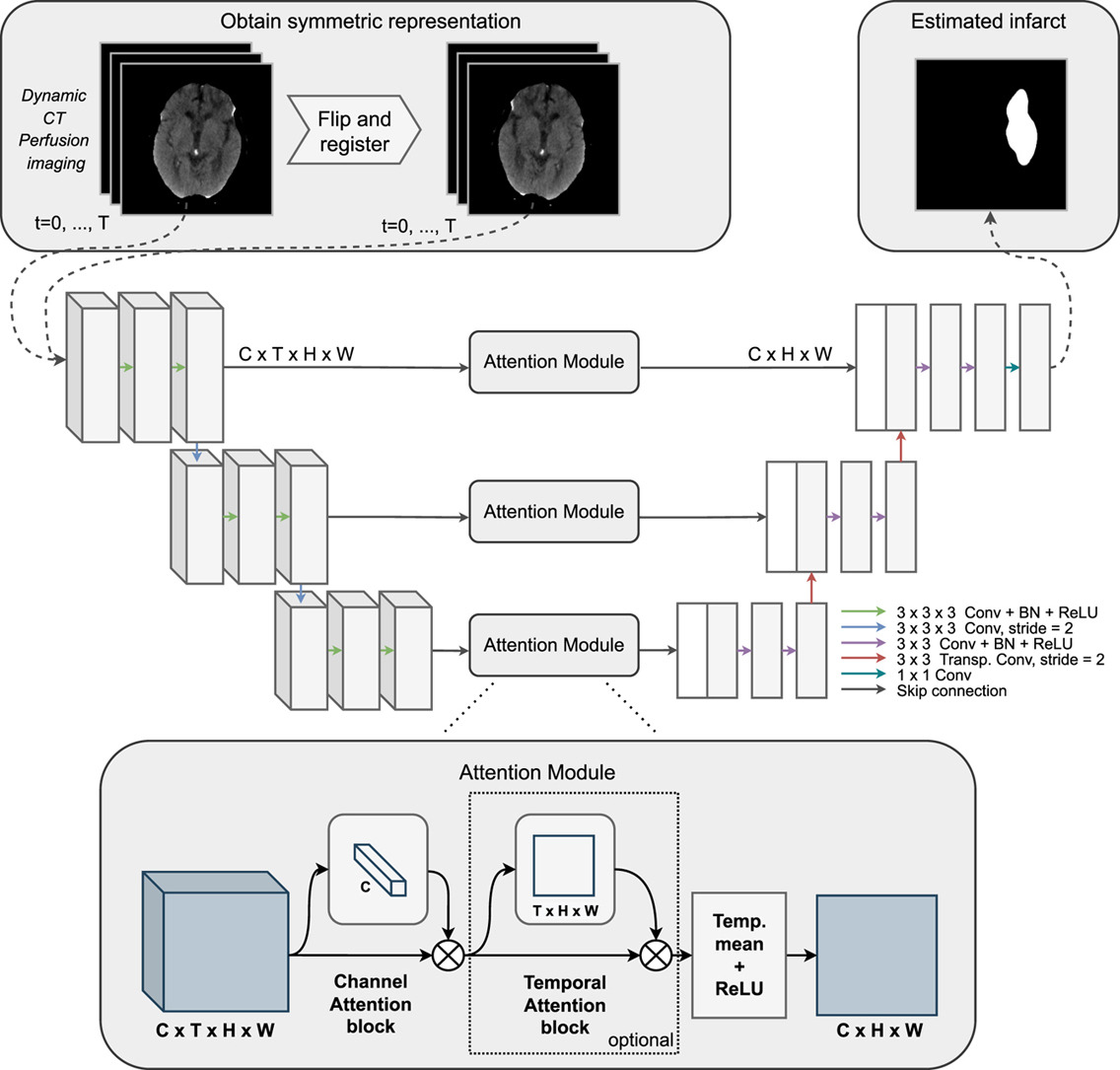}
    \caption{ PerfU-Net architecture \cite{de2023perfu}. }
    \label{fig:perfunet}
\end{figure}

\begin{figure}
    \centering
    \includegraphics[width=.4\textwidth]{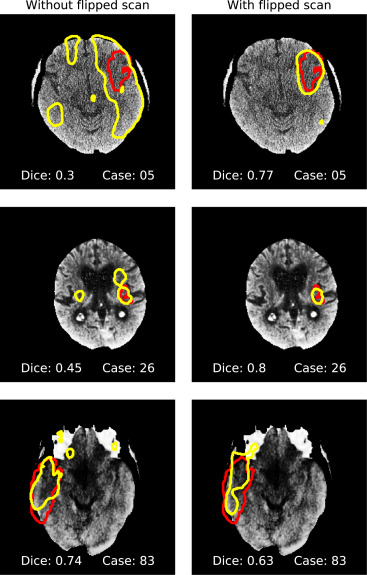}
    \caption{Qualitative analysis of PerfU-Net performance for three cases from ISLES 2018 dataset with and without flipped representation \cite{de2023perfu}. Yellow lines indicate the predicted segmentation mask and red lines indicate the ground truth. }
    \label{fig:perfunet-isles}
\end{figure}

PerfU-Net \cite{de2023perfu} was introduced for stroke segmentation from CT images. This architecture incorporated an attention module placed in the skip connections, with two variations tested: one featuring channel attention and the other incorporating both channel attention and temporal attention. The training process utilized the generalized Dice loss \cite{sudre2017generalised} as the loss function. PerfU-Net achieved a mean Dice coefficient of 0.564 when evaluated on the ISLES 2018 dataset, utilizing 32 frames as the input to the model and employing channel attention as the attention module. Refer to Figure \ref{fig:perfunet} for an illustration of the proposed pipeline in PerfU-Net, and refer to Figure \ref{fig:perfunet-isles} for a qualitative analysis of performance using the ISLES 2018 dataset under two conditions: with and without flipped scans. In their observations, they noted that the use of flipped scans contributed to a reduction in the number of false positives.

\begin{table*}
    \centering
    \caption{Performance Analysis of CNN-Based Approaches for Stroke Segmentation. ${}^{\ast}$ indicates the median. }
    \resizebox{\textwidth}{!}{
    \begin{tabular}{|l|l|l|c|c|c|c|c|}
        \hline
        \multirow{2}{*}{\hfil Reference} & \multirow{2}{*}{\hfil Year} & \multirow{2}{*}{\hfil Dataset} & \multicolumn{5}{c|}{\text{Performance}} \\
        \hhline{|~ ~ ~|-----|}
        & & & \text{Dice} & \text{Sensitivity/Recall} & \text{Precision} & \text{HD} & \text{Others}\\
        \hline
    \cite{praveen2018ischemic} & 2018 & ISLES 2015-SISS training dataset & $0.943 \pm 0.057$ & $0.924 \pm 0.072$ & $0.968 \pm 0.074$ & - & -\\
    \hline 
    \cite{liu2019efficient}& 2019 & \makecell[tl]{ISLES 2015-SISS \\ ISLES 2015-SPES} & \makecell[tc]{$0.57 \pm 0.29$ \\ $0.76 \pm 0.11$} & \makecell[tc]{- \\- }& \makecell[tc]{- \\-} &  \makecell[tc]{$43.02 \pm 30.48$ \\$36.93 \pm 25.42$}&\makecell[tc]{ASSD: $8.22 \pm 16.54$ \\ ASSD: $1.79 \pm 0.54$} \\
    \hline
     \cite{clerigues2020acute} & 2020 & \makecell[tl]{ISLES 2015-SISS \\ ISLES 2015-SPES} & \makecell[tc]{$0.59 \pm 0.31$ \\ $0.84 \pm 0.10$}& \makecell[tc]{$0.60 \pm 0.30$ \\ $0.89 \pm 0.06$}& \makecell[tc]{ $0.65 \pm 0.35$ \\ $0.82 \pm 0.15$} & \makecell[tc]{$34.7 \pm 28.9$ \\ $20.7 \pm 13.9$} & \makecell[tc]{- \\ -}\\
    \hline
     \cite{zhang2020ischemic} & 2020 & ISLES 2015-SSIS & $0.622$& $0.541$& - & - & IoU: $0.4514$ \\
    \hline
    \cite{bal2023robust} & 2023 & \makecell[tl]{ISLES 2015-SISS \\ ISLES 2015-SPES} & \makecell[tc]{$0.87 \pm 0.10$ \\ $0.90 \pm 0.08$} & \makecell[tc]{$0.86 \pm 0.09$ \\ $0.89 \pm 0.08$} & \makecell[tc]{ $0.88 \pm 0.10$ \\   $0.90 \pm 0.07$}  & \makecell[tc]{- \\ -}& \makecell[tc]{- \\ -}   \\
    \hline
    \cite{pereira2019adaptive} & 2019 & \makecell[tl]{ISLES 2015-SPES \\ ISLES 2017}& \makecell[tc]{$0.82 \pm 0.09$ \\ $0.34 \pm 0.20$ } & \makecell[tc]{- \\ $0.55 \pm 0.30$} & \makecell[tc]{- \\ $0.36 \pm 0.25$} & \makecell[tc]{- \\ -} & \makecell[tc]{ASSD: $1.27 \pm 0.72$ \\ -} \\
    \hline
    \cite{islam2018class} & 2018 & ISLES 2017 & $0.29 \pm 0.2$& $0.6 \pm 0.24$ & $0.35 \pm 0.23$ & $46.9 \pm 12.8$ & -\\
    \hline
    \cite{lucas2018multi} & 2018 & ISLES 2017 & $0.35$ & $0.35$ & $0.52$ & $21.48$ & ASSD: $3.45$\\
    \hline
    \cite{hu2020brain} & 2020 & ISLES 2017 & $0.30 \pm 0.22$ & $0.43 \pm 0.27$ & $0.35 \pm 0.27$ & - & -\\
    \hline
    \cite{abulnaga2019ischemic} & 2019 & ISLES 2018 & $0.54 \pm 0.09$ & - & - & - & -\\
    \hline
    \cite{clerigues2019acute} & 2019 & ISLES 2018 testing dataset &  $0.49 \pm 0.31$ & $0.57 \pm 0.35$ &  $0.51 \pm 0.36$ & $11.3 \pm 31.6$ & -\\
    \hline
    \cite{rubin2019ct} & 2019 & ISLES 2018 &  $0.54 \pm 0.23$ & $0.63 \pm 0.25$ & $0.56 \pm 0.25$ & $27.88 \pm 21.00$ & - \\
    \hline
    \cite{liu2022pool} & 2022 & ISLES 2018 & $0.565 \pm 0.205$ & $0.565 \pm 0.218$ & $0.678 \pm 0.223$ & $21.14 \pm 13.61$& - \\
    \hline
    \cite{de2023perfu} & 2023 & ISLES 2018 & $0.564 \pm 0.009$ & $0.644 \pm 0.008$ & $0.565 \pm 0.019$ & $22.3 \pm 1.8$ & AVD: $9.9 \pm 0.4$\\
    \hline
    \cite{qi2019x} & 2019 & ATLAS v1.2 & $0.4867$ & $0.4752$& $0.6000$& - & IoU: $0.3723$ \\
    \hline
    \cite{zhou2019d} & 2019 & ATLAS v1.2 & $0.534 \pm 0.276$ & $0.524 \pm 0.291$ & $0.6331 \pm 0.2958$& - & -\\
    \hline
    \cite{liu2019msdf} & 2019 & ATLAS v1.2 & $0.5578 $ & $0.8291 ^{\ast}$ & - & - & VOE: $0.1403 ^{\ast}$ \\
    \hline
    \cite{yang2019clci} & 2019 & ATLAS v1.2 & $0.581$ & $0.581$ & $0.649$ & - & \makecell[tc]{VOE: $54.6$ \\ RVD: $25.4$}\\
    \hline
    \cite{basak2020f} & 2020 & ATLAS v1.2 & $0.535 \pm 0.276$& $0.523 \pm 0.293$ & $0.634 \pm 0.287$& - & - \\
    \hline
    \cite{hui2020partitioning} & 2020 & ATLAS v1.2 & $0.593$& $0.62$& $0.691$& - & - \\
     \hline
     \cite{tomita2020automatic} & 2020 & ATLAS v1.2 & $0.64$ & - & $0.62$ &  $20.4$& ASSD: $3.6$\\
     \hline
     \cite{yu2023fan} & 2023 & ATLAS v1.2 & $0.559 \pm 0.180$& $0.576 \pm 0.162$ & - & - & F1: $0.545 \pm 0.162$\\
    \hline
    \cite{yu2023san} & 2023 & ATLAS v1.2 & $0.571 \pm 0.195$ & $0.597 \pm 0.158$ & - & - & F1: $0.562 \pm 0.192$\\
    \hline
    \cite{huo2022mapping} & 2022 & ATLAS v2.0 training dataset  & $0.646 \pm 0.270$ & - & - & $21.51 \pm 26.82$ & \makecell[tc]{VD: $5729 \pm 11565$ \\  SLC: $3.382 \pm 6.786$}\\
    \hline
    \cite{chalcroft2023large} &  2023 & \makecell[tl]{ISLES 2022 \\ ATLAS v2.0} & \makecell[tc]{$0.693$ \\ $0.678 ^ {\ast}$}& \makecell[tc]{- \\ -}& \makecell[tc]{- \\ -}& \makecell[tc]{- \\ -} & \makecell[tc]{F1: $0.657$ \\ F1: $0.474^ {\ast}$}\\
    \hline
    \cite{liang2021symmetry} & 2021 & AISD & $0.5784$ & $0.5880$& $0.6597$ & - & F1: $0.6218$ \\
    \hline
     \cite{ni2022asymmetry} & 2022 & AISD & 0.5245 & - & - & 39.18 & - \\
    \hline
    \end{tabular} 
    }
    \label{Tab:cnn}
\end{table*}

   \subsection{Transformer-Based Architectures for Stroke Segmentation}
Vision Transformers have been employed in recent years for stroke segmentation, leveraging their capabilities, especially when combined with CNNs to capture local and global information from the input data. Table \ref{Tab:transformer} summarizes the performance of selected Transformer-based methods for stroke segmentation across various datasets. \cite{de2021transformers} proposed a hybrid Transformer-CNN pipeline for ischemic stroke infarct segmentation from CT perfusion scans. They considered the axial slices of 3D images as temporal information and incorporated the flipped and registered form of each slice as an additional channel in the input to exploit the brain's bilateral quasi-symmetry property. The spatio-temporal data were fed into a Transformer block consisting of the Linformer \cite{wang2020linformer} backbone to generate an attention map representing the probability of infarction. Subsequently, this attention map, along with the source data, was input into a traditional U-Net to produce the final segmentation. The Transformer part was trained using cross-entropy loss, while the U-Net was trained using the generalized Dice loss function.

UCATR \cite{luo2021ucatr} was proposed for acute ischemic stroke segmentation from non-contrast CT images. A Transformer-based block succeeded the CNN encoder in the bottleneck, and irrelevant information was filtered by employing Multi-Head Cross-Attention modules in the skip connections. The proposed network was evaluated on a clinical dataset containing information from 11 patients, averaging 95 slices per patient, and it outperformed U-Net, Attention U-Net, and TransUNet by achieving a mean Dice coefficient of 0.7358. UTransNet \cite{feng2022utransnet} employed a novel module (CT block) consisting of two convolutional layers and a Transformer module to leverage the advantages of both CNNs and Transformers. To mitigate computational complexity within the self-attention mechanism, the PVT v2 Transformer \cite{wang2022pvt} was incorporated. In evaluations conducted on the ATLAS dataset, UTransNet achieved superior results compared to other Transformer-based methods, including TransUNet, SwinUNet \cite{cao2022swinunet}, and UCTransNet \cite{wang2022uctransnet}.

The Multi-Encoder Transformer (METrans) \cite{wang2022metrans} was proposed as a novel architecture, incorporating a methodology involving additional encoding modules. These modules served the purpose of extracting abstract features at distinct stages of the primary encoder path. The ensuing step involved the fusion of the multi-scale extracted features. Following each convolutional module within the encoder, Convolutional Block Attention Modules (CBAM) \cite{woo2018cbam} were employed to harness both channel attention and spatial-channel attention. Furthermore, a Transformer-based block was integrated into the bottleneck to facilitate the extraction of global features.

Within the architecture of STHarDNet \cite{gu2022sthardnet}, HarDNet blocks \cite{chao2019hardnet} were employed in both the encoder and decoder paths. Notably, the Swin Transformer was incorporated exclusively in the initial layer of the skip connection, while the subsequent layers adhered to a conventional, plain structure. The performance of this network surpassed that of numerous CNN-based and Transformer-based counterparts when evaluated on the ATLAS dataset. LLRHNet \cite{liu2022llrhnet} implemented a dual-path approach for feature encoding, wherein the initial path utilized CNN layers to extract local information, and the subsequent path employed a Transformer-based block for encoding global features. The features extracted from these two paths were concatenated, and a final prediction was generated through a CNN decoder. To enhance information transfer from the CNN encoder to the decoder, the model integrated multi-level feature fusion skip connections, a departure from conventional skip connection methods. Evaluation of the LLRHNet on a clinical dataset for ischemic stroke segmentation demonstrated its superior performance by achieving a mean Dice coefficient of 0.791. 

In \cite{wu2022multi} an architecture consisting of three main elements was proposed. First, the Patch Partition Block (PPB) was employed to encode the image as a patch sequence, simultaneously reducing the number of parameters. Second, the Multi-scale Long-Range Interactive and Regional Attention (MLiRA) mechanism served as the encoder, comprising multiple subsampling Transformers (STR) followed by convolutional blocks. Within STR, subsampling Multi-head Interactive Self-Attention mechanisms were utilized to capture dimensional interactive attention. Moreover, STR exhibited flexibility in adjusting input resolution to attain global information at various spatial resolutions. Third, the Feature Interpolation Path (FIP) was utilized as the decoder, facilitating the recovery of encoded features to the original image resolution.

In \cite{marcus2023concurrent} a multi-task Transformer-based network for age estimation and segmentation of ischemic lesions from CT images was proposed. Their architecture was rooted in the DETR architecture \cite{carion2020end} with certain modifications. The primary components of the proposed network included: 1) a CNN encoder consisting of four ResNeXt \cite{xie2017aggregated} blocks to generate an activation map; 2) a Transformer encoder-decoder, commencing with a pyramid pooling module \cite{zhao2017pyramid} to augment the receptive field, followed by a Transformer block using the gated positional self-attention mechanism \cite{d2021convit}; 3) heads for age estimation and bounding box predictions; and 4) a CNN decoder serving as the segmentation head. Based on evaluations of their proposed pipeline on a large clinical dataset consisting of 776 CT images collected from two medical centers, they reached a mean Dice coefficient of 0.382. For evaluation of the generalizability of the trained network on unseen data, they also utilized the ISLES 2018 dataset as the test dataset and reached a 0.203 mean Dice coefficient.

\cite{zhang2023efficient} introduced a pipeline designed to address the efficient processing of 3D image data to preserve volumetric information. Their proposed architecture, named PDSwin, leverages the Swin Transformer with a pyramidal downsampling approach, spatially downsampling 2D slices. Additionally, the authors addressed the shift domain issue arising from diverse image acquisition centers by proposing a cluster-based domain adversarial algorithm. In \cite{xu2023combining} a U-shaped network was employed to segment stroke infarct from CT scans. In the CNN-based encoder path, multiple CBAM blocks were incorporated. The features extracted from two scales of the encoder path were flattened and inputted into a Transformer module, encompassing several deformable Transformer layers utilizing deformable self-attention \cite{zhu2020deformable}. 

\cite{soh2023hut} introduced a Hybrid UNet and Transformer (HUT) network, comprising two parallel stages: a UNet and a Transformer block. The input to the Transformer block consisted of intermediate features extracted by the CNN encoder of the UNet. The output of the Transformer block was then fused with the extracted features from the encoder path in the skip connections at two different scales. In SAMIHS \cite{wang2023samihs} a parameter-efficient fine-tuning strategy to the Segment Anything Model (SAM) \cite{kirillov2023segment} model was applied to segment hemorrhagic stroke. To improve segmentation results, they utilized a combination of the binary cross-entropy loss and a boundary-sensitive loss.

Some research endeavours have been pursued to improve the realism of predicted segmentation mask boundaries in stroke nature. One such approach was TransRender \cite{wu2023transrender}, which was proposed to address the issue of overly smooth boundaries. It achieved this by adaptively selecting specific points for computing the boundary features in a point-based rendering manner, intending to enhance the fidelity of the boundary estimation. The hierarchical Transformer-based encoder path captured global information across multiple scales, with additional parallel CNN blocks employed to capture local information. Both local and global features were then provided as input to multiple render modules. These modules, by selecting specific uncertain points and extracting feature representations for those points, facilitated the re-prediction of these uncertain points as boundary points.

\begin{figure*}
    \centering
    \includegraphics[width=.9\textwidth]{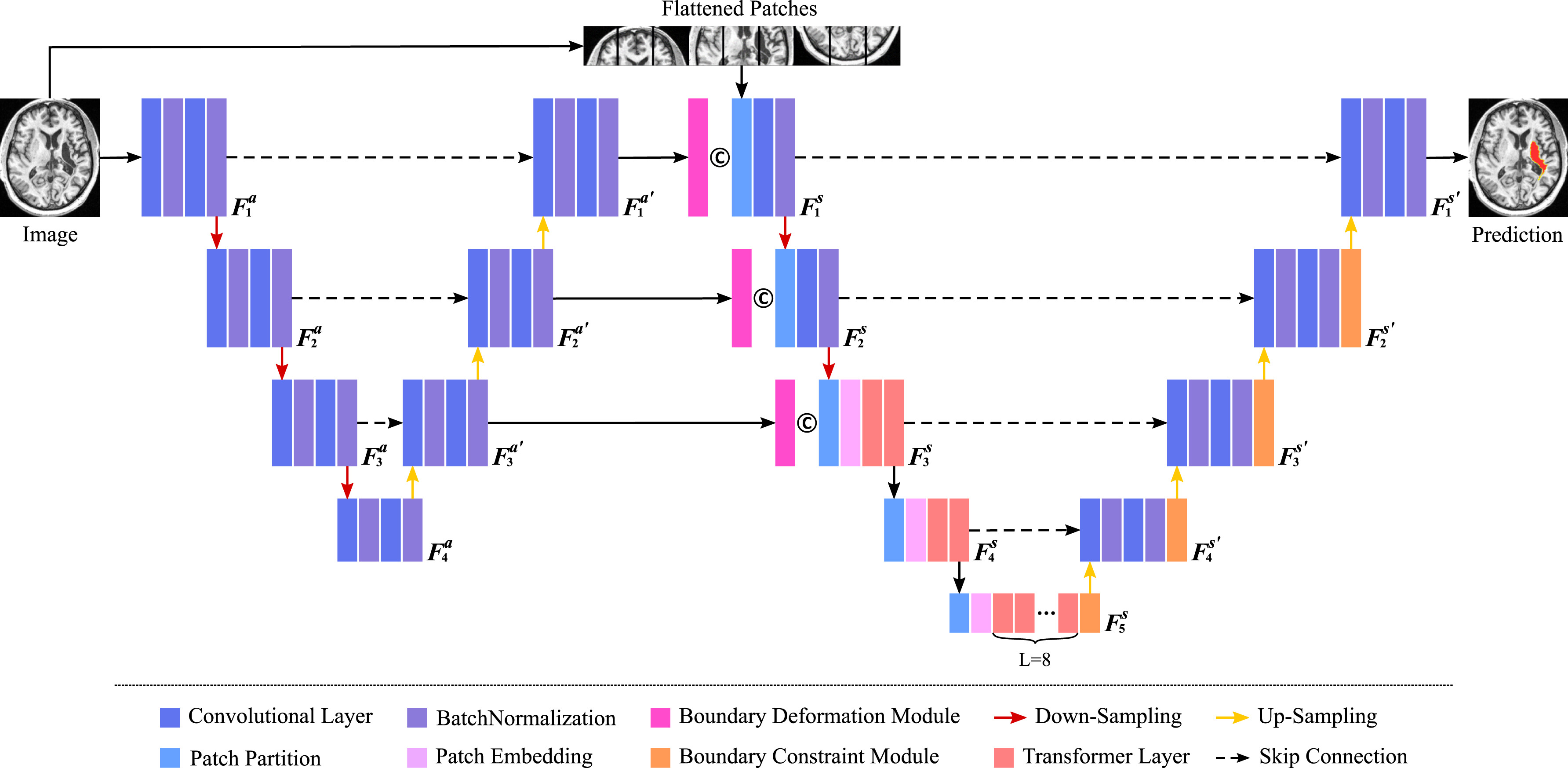}
    \caption{W-Net architecture \cite{wu2023w}.}
    \label{fig:wnet}
\end{figure*}

\begin{figure}
    \centering
    \includegraphics[width=.8\textwidth]{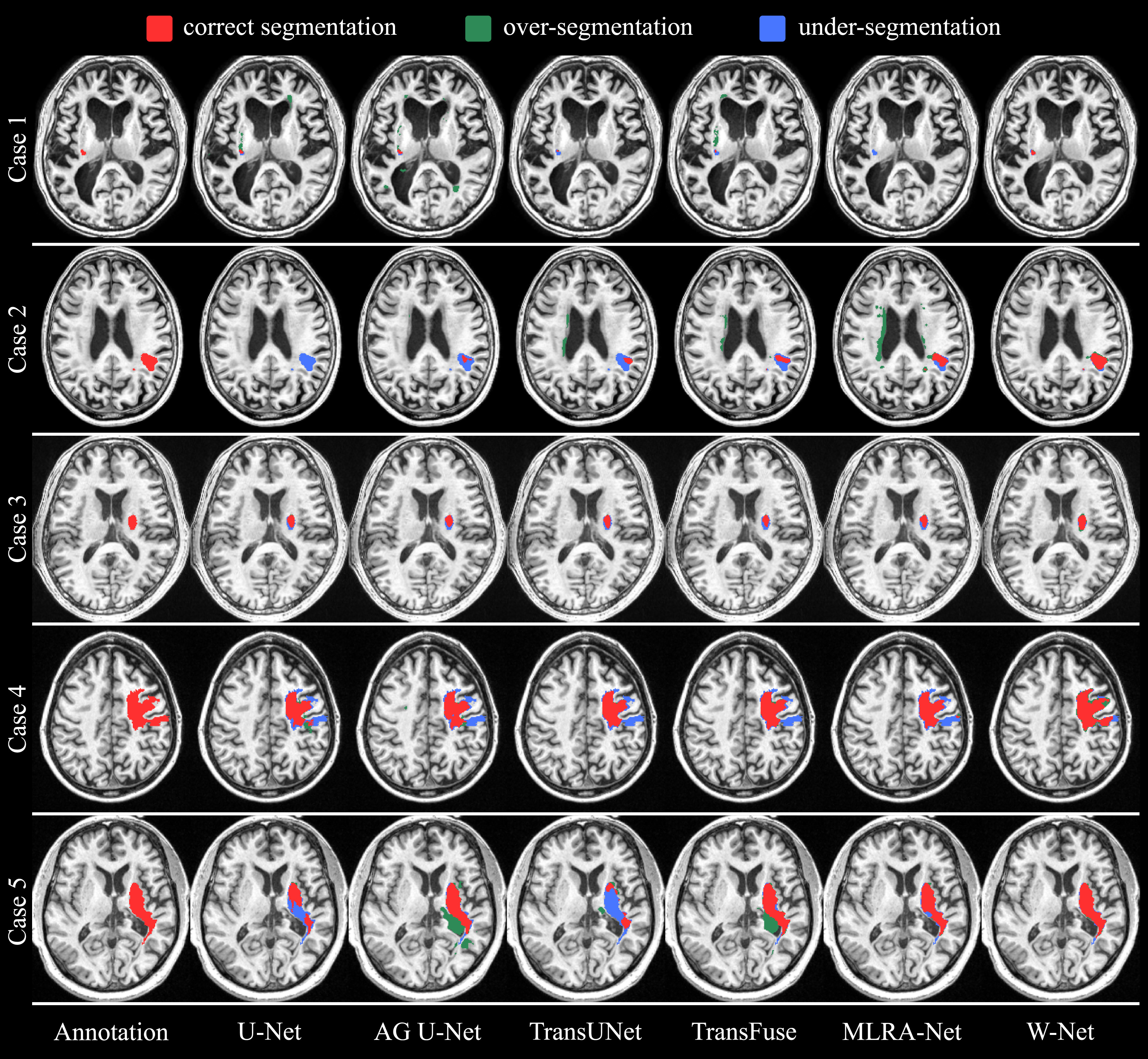}
    \caption{Qualitative analysis of the W-Net performance compared to other five benchmarks using ATLAS dataset \cite{wu2023w}.}
    \label{fig:wnet-atlas}
\end{figure}

\begin{figure}
    \centering
    \includegraphics[width=.8\textwidth]{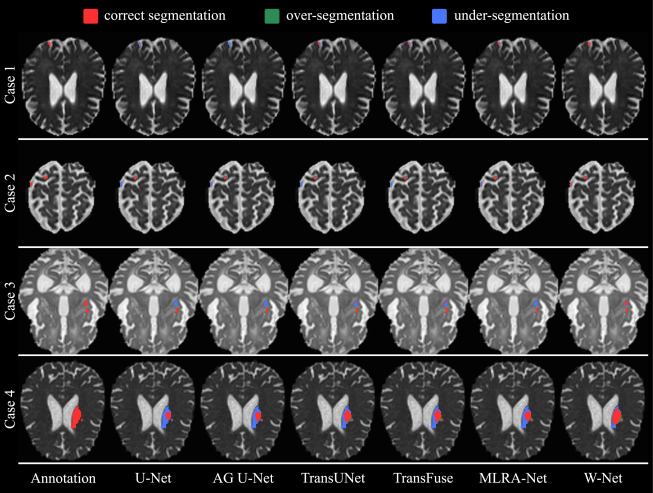}
    \caption{Qualitative analysis of the W-Net performance compared to other five benchmarks using ISLES 2022 dataset \cite{wu2023w}.}
    \label{fig:wnet-isles}
\end{figure}

Another approach to enhance boundary estimation accuracy was W-Net \cite{wu2023w}, which introduced a Boundary Deformation Module (BDM) and a Boundary Constraint Module (BCM) to address fuzzy boundaries. W-Net integrated both a CNN network and a Transformer-based network as backbone networks. Initially, a U-shaped CNN network was employed for coarse segmentation, leveraging the advantages of CNNs to extract local features. In the decoder path, features of different scales were inputted into proposed BDM blocks for further optimization through iterative boundary deformation, correcting the initially predicted boundaries using circular convolutions. The second stage of W-Net consisted of a Transformer-based U-shaped network. The output of the BDM blocks was fused with the encoder features at multiple levels. The decoder utilized BCM blocks to refine the encoded global features by constraining the boundary curves, employing several parallel dilated convolution layers. Refer to Figure \ref{fig:wnet} for an illustration of the W-Net architecture, and Figures \ref{fig:wnet-atlas} and \ref{fig:wnet-isles} for qualitative analyses on the ATLAS and ISLES 2022 datasets, respectively.

\begin{table*}
    \centering
    \caption{Performance Analysis of Transformer-Based Approaches for Stroke Segmentation.  }
   \resizebox{\textwidth}{!}{
    \begin{tabular}{|l|l|l|c|c|c|c|c|}
        \hline
        \multirow{2}{*}{\hfil Reference} &  \multirow{2}{*}{\hfil Year} & \multirow{2}{*}{\hfil Dataset} & \multicolumn{5}{c|}{\text{Performance}} \\
        \hhline{|~ ~ ~|-----|}
        & & & \text{Dice} & \text{Sensitivity/Recall} & \text{Precision} & \text{HD} & \text{Others}\\
        \hline
        \cite{de2021transformers} & 2021 & ISLES 2018 test dataset & $0.42$ & $0.53$ & $0.44$ & - & - \\
        \hline
        \cite{xu2023combining} & 2023 & ISLES 2018 & $0.4667$ & $0.4724$ & $0.5888$ & - & F1: $0.5242$\\
        \hline
        \cite{wang2022metrans} & 2022 & \makecell[tl]{ISLES 2018 \\ ATLAS v1.2} & \makecell[tc]{$0.67$ \\ $0.931$} & \makecell[tc]{$0.641$ \\ $0.91$} & \makecell[tc]{ $0.72$ \\  $0.94$}& \makecell[tc]{- \\ -} &  \makecell[tc]{- \\ -}\\
        \hline
        \cite{wu2023transrender} & 2023 & \makecell[tl]{ISLES 2022 \\ ATLAS v1.2 } & \makecell[tc]{$0.8537$ \\ $0.5979$} & \makecell[tc]{$0.8394$ \\ $0.6808$} & \makecell[tc]{$0.8648$ \\ $0.6391$} & \makecell[tc]{$27.60$ \\ $33.98$} & \makecell[tc]{F2: $0.8487$ \\ F2: $0.5938$ }\\
         \hline
         \cite{wu2023w} & 2023 & \makecell[tl]{ISLES 2022 \\ ATLAS v1.2 } & \makecell[tc]{$0.8560$ \\ $0.6176$} & \makecell[tc]{$0.8539$ \\ $0.6868$} & \makecell[tc]{$0.8834$ \\ $0.6286$} & \makecell[tc]{ $27.34$ \\ $32.47$} & \makecell[tc]{F2: $0.8529$ \\ F2: $0.6460$} \\
         \hline
        \cite{feng2022utransnet} & 2022 & ATLAS v1.2 & $0.8597$ & - & - &- & \makecell[tc]{F1: $0.8494$ \\ IoU: $0.8251$}\\
        \hline
        \cite{gu2022sthardnet} & 2022 & ATLAS v1.2 & $0.5547$ & $0.5286$ & $0.6764$ & - & IoU: $0.4184$\\
        \hline
        \cite{wu2022multi} & 2022 & ATLAS v1.2 & $0.6119$ & $0.6765$ & $0.6330$ & $13.49$ & F2: $0.6376$\\
        \hline
        \cite{soh2023hut} & 2023 & ATLAS v1.2 & $0.737 \pm 0.127$ & $0.706 \pm 0.153$ & $0.825 \pm 0.172$ & $10.335 \pm 10.074$ & IoU: $0.598 \pm  0.114$\\
        \hline
        \cite{zhang2023efficient} & 2023 & ATLAS v2.0 & $0.6273$ & - & - & - & -\\
        \hline
        \cite{wang2023samihs} & 2023 & \makecell[tl]{IHS \\ INSTANCE} & \makecell[tc]{$0.6977$ \\ $0.7652$} &  \makecell[tc]{ - \\ -} & \makecell[tc]{- \\ -} & \makecell[tc]{$3.31$ \\ $3.71$} &  \makecell[tc]{ - \\ -} \\
        \hline  
    \end{tabular}
    }
   \label{Tab:transformer}
\end{table*}

\section{Open Challenges and Future Directions}

Current solutions for stroke segmentation, whether they employ CNN networks or Transformer-based architectures, have shown less satisfactory results compared to tasks such as tumor segmentation \cite{ranjbarzadeh2023brain, liu2023deep}. The underlying cause of this suboptimal performance is attributed to various factors, including high variability in the location, number, size, and pattern of the infarct. Furthermore, the intensity differences resulting from the varied imaging vendors and stroke ages pose a significant challenge for automated algorithms. The proposed methods must effectively distinguish between healthy and infarcted regions of the brain while accommodating the diverse variability introduced by different medical imaging systems and inherent stroke features.
 
An additional crucial feature of proposed methods in stroke segmentation is their generalizability to unseen data from different vendors, making them applicable. Current methods often exhibit a lack of generalization, with testing on unseen data acquired from diverse centers leading to significantly lower performance. It is imperative to investigate and improve the robustness of these methods to handle unseen data effectively. Exploring domain adaptation methods could prove beneficial in achieving this objective.
 
Another inherent characteristic of stroke infarcts is their ability to affect various parts of the brain at different stages, exhibiting different sizes in different locations. It is crucial for segmentation methods to accurately handle multi-instance infarcts of varying sizes. Despite the high values of the Dice coefficient suggesting good overall performance, instance-wise measurements often yield lower results \cite{kofler2023panoptica}. This discrepancy is primarily due to the neglect of small infarcts in the presence of larger ones when calculating commonly used metrics. The currently proposed architectures encounter challenges in detecting and segmenting small infarcts, resulting from information loss within the deeper layers of the networks designed to represent abstract features. Thoughtful improvements are necessary to adapt proposed pipelines for the segmentation of small infarcts.

Transformers are widely employed in medical image segmentation due to their ability to represent global information and capture long-range dependencies, providing a robust representation of shape-based features for segmentation. Given the high variability in stroke infarct shape, location, and pattern, the incorporation of texture information derived from CNNs appears to be more beneficial. Hybrid CNN-Transformer architectures address this challenge by combining texture-based and global information. However, a careful selection of how to employ Transformer blocks is necessary to fully leverage their advantages for the integration of texture information alongside global features to improve the effectiveness of the stroke segmentation methodologies.

A notable challenge in the effective application of Transformers stems from their dependence on large datasets, which becomes especially pronounced in the medical field. The limited availability of labeled data, coupled with difficulties in acquiring annotations and privacy considerations, imposes constraints on the accessibility of medical data. Although pre-training on alternative datasets, including natural images, might be considered, it is suboptimal due to the inherent domain shift. Medical images often have high dimensions, resulting in a large number of parameters that demand significant computational resources. The increased parameter count, combined with a limited dataset, can lead to overfitting issues. Various strategies, such as slicing 3D data from different anatomical planes or adopting patch-based data inputting, have been attempted previously. However, these approaches can result in the loss of valuable information. Exploring realistic data augmentation techniques and optimizing data input methods are crucial avenues to investigate in order to mitigate these challenges.

To establish effective pipelines for stroke segmentation, it is crucial to incorporate additional characteristics, such as privacy-preserving algorithms \cite{sheller2020federated, li2021fedbn, li2019convergence}, and embrace explainable artificial intelligence (XAI) \cite{alicioglu2022survey, mondal2021xvitcos, singh2020explainable} as integral components of trustworthy AI \cite{liu2022trustworthy}. In recent years, researchers have dedicated their efforts to interpret deep learning-based models, which were previously considered black boxes. Understanding the decision-making process of Transformers can enhance predictions and facilitate their use in aiding decision-making for medical diagnoses. Additionally, exploring privacy-preserving algorithms is imperative. This investigation aims to provide a platform that can be utilized in different centers, allowing the sharing of knowledge derived from diverse datasets without compromising the privacy of medical information. This approach aims to continually enhance the performance of the underlying pipeline.

\section{Discussion and Conclusion}
In this paper, we present a comprehensive review of Transformer-based architectures for segmenting stroke infarcts from MRI and CT images. We begin by offering preliminary information on the concepts of self-attention, vision Transformers, and several benchmark Transformer-based networks designed for medical image segmentation. Following this, we delve into details about available datasets for stroke segmentation, encompassing both ischemic and hemorrhagic strokes for both MRI and CT modalities. Subsequently, we discussed commonly used metrics for evaluating segmentation performance and conducted a literature review on stroke segmentation using deep learning methods. Given that a significant portion of previous research has been conducted using CNNs, we selectively extracted and summarized the key ideas with superior performance. Specifically focusing on Transformer-based architectures, we conducted a review of 15 papers, all employing hybrid CNN-Transformer architectures. For each paper, we offered a high-level abstraction of the core techniques utilized in these networks. Additionally, we presented comparison tables for quantitative evaluations of performance, considering both CNN-based and Transformer-based networks. Finally, we outlined the unsolved challenges associated with stroke segmentation and suggested potential avenues for future research directions.

\section*{Acknowledgment}
The findings presented in this paper have emerged from a project funded by the Qatar Japan Research Collaboration Research Program under grant number M-QJRC-2023-313. The authors extend their sincere gratitude to Marubeni and Qatar University for their consistent and generous support.

\bibliography{refs}

\end{document}